\documentclass[letterpaper,journal]{IEEEtran}
\usepackage{comment}
\usepackage{amsmath,amsfonts}
\usepackage{algorithmic}
\usepackage{algorithm}
\usepackage{mathtools}
\usepackage{amssymb}
 \usepackage{balance}
\usepackage{subfigure}
\usepackage[caption=false,font=normalsize,labelfont=sf,textfont=sf]{subfig}
\usepackage{amsthm}
\usepackage{array}
\usepackage{textcomp}
\usepackage{stfloats}
\usepackage{url}
\usepackage{verbatim}
\usepackage{graphicx}

\newtheorem{lemma}{Proposition}[section]

\usepackage{cite}
\usepackage[T1]{fontenc}
\usepackage{soul}
\usepackage{xcolor}

\DeclarePairedDelimiter\abs{\lvert}{\rvert}

\usepackage{xcolor,cite,etoolbox}

\makeatletter 
\pretocmd\@bibitem{\color{black}\csname keycolor#1\endcsname}{}{\fail}
\newcommand\citecolor[1]{\@namedef{keycolor#1}{\color{red}}}
\makeatother

\begin{document}


\title{\textcolor{black}{On the Performance of THz Wireless Systems over $\alpha$-$\mathcal{F}$ Channels with Beam Misalignment, Mobility and
Hardware Impairments}}


\author{
 Wamberto J. L. Queiroz, Higo T. P. Silva, Hugerles S. Silva,~\IEEEmembership{Senior Member,~IEEE} and Alexandros-Apostolos A. Boulogeorgos,~\IEEEmembership{Senior Member,~IEEE}
 \\
\thanks{

W. J. L. Queiroz is with the Department of Electrical Engineering, Federal University of Campina Grande, Campina Grande, Paraíba, Brazil (e-mail: wamberto@dee.ufcg.edu.br).

H. T. P. Silva and H. S. Silva are with the Electrical Engineering Department, University of Brasília (UnB), Federal District, Brazil, (e-mail: higo.silva@unb.br, hugerles.silva@unb.br).

A.-A. A. Boulogeorgos is with the Department of Electrical and Computer
Engineering, University of Western Macedonia, ZEP Area, 50100 Kozani,
Greece, (e-mail: aboulogeorgos@uowm.gr).


}}


\markboth{Submitted to IEEE Transactions on Vehicular Technology, April~2026}{}%


\maketitle

\begin{abstract}
This paper investigates the performance of terahertz~(THz) wireless systems over the $\alpha$-$\mathcal{F}$ fading channels with beam misalignment, mobility \textcolor{black}{and hardware impairments}.
New expressions are derived for the probability density, cumulative distribution, and 
higher-order moments of the instantaneous signal-to-noise ratio~(SNR).
Building upon the aforementioned expressions, we extract novel formulas for the outage probability~(OP), average symbol error probability, and average channel capacity. 
\textcolor{black}{Asymptotic expressions are also derived, providing useful insights into system performance in the high-SNR regime. Furthermore, an upper bound on the capacity metric is obtained.} Monte Carlo simulation results are presented to validate the developed analytical framework.
\end{abstract}


\section{Introduction}

\IEEEPARstart{T}{erahertz} (THz) wireless systems have emerged as key enablers for future networks, offering extremely ultra-wide bandwidths and high data rates~\cite{Thomas2025}.
Over the years, the growing interest in THz wireless systems for sixth-generation (6G) networks underscores the importance of developing accurate channel models that jointly characterize the various phenomena that affect signal propagation~\cite{Thomas2025}. THz links are simultaneously influenced by multipath fading, shadowing, beam misalignment, \textcolor{black}{hardware impairments}, mobility, path loss, and atmospheric molecular attenuation,
whose combined impact must be properly modeled to enable realistic system-level performance analysis. For prospective 6G application scenarios, the use of highly directional beams is essential to overcome severe path losses and molecular absorptions at THz frequencies.
However, due to mobility, a misalignment between the transmitting and receiving antennas may occur, which not only degrades the link quality, but can also lead to severe communication outages~\cite{Dabiri}.
Scanning the open technical literature, several contributions can be identified that focus on the problem of beam misalignment, \textcolor{black}{hardware impairments} and/or mobility.

\color{black}

In~\cite{Chen}, using stochastic geometry, a system level beam misalignment model and a throughput analysis for THz networks was provided by considering the joint impact of high-directionality, mobility, blockage, and molecular absorption.
In~\cite{Pattaramalai}, the probability density function (PDF) and cumulative distribution function~(CDF) of the received power under dynamic heterogeneous wireless networks, considering fading channels and mobility, were articulated.
Based on these expressions, the outage probability (OP) and average bit error rate (ABER) were deduced to quantify the performance of the mobile system. In the mentioned work, the authors adopted the $\eta$-$\mu$ distribution and the random waypoint model~(RWP) to characterize the fading and mobility, respectively.

The performance of a mobile wireless network over the Fisher–Snedecor $\cal{F}$ composite fading channel was presented in~\cite{Shawaqfeh}, where a multi-antenna base station employs maximum-ratio transmission to send information to a single-antenna mobile receiver.
In~\cite{Shawaqfeh}, receiver mobility was modeled using RWP.
An analysis was performed by the authors in~\cite{OsamahPoint}, considering 
THz wireless systems under $\alpha$-$\mu$ channels with pointing
errors. Several expressions were derived for first-order statistics and the performance was assessed by means of OP, channel capacity, and symbol error rate~(SER) metrics.
A conventional and a non-conventional system with a robust autoencoder was evaluated in~\cite{Pereira}, considering mobility, path loss, and composite $\alpha$-$\mu$/gamma fading. The authors derived new and closed-form statistics expressions, and the performance analysis extracted the OP, average symbol error probability (ASEP) and channel capacity metrics.
THz systems, considering the fading characterized by mixture gamma jointly with the beam misalignment effect, were studied in~\cite{Jemaa}.
Closed-form expressions were derived for bit error probability, OP and ergodic capacity, as well as an asymptotic analysis was performed at high
signal-to-noise ratio~(SNR), under different configurations. 

\textcolor{black}{Several works have also been reported in the literature addressing the effects of transceiver hardware impairments~\cite{Bjornson,Boulogeorgos,Bhardwaj}. This is motivated since practical transceivers exhibit hardware imperfections that introduce distortions, thereby degrading the system performance~\cite{Bjornson}.
An analytical framework for evaluating the joint impact of misalignment fading, hardware impairments and multipath fading characterized by the $\alpha$-$\mu$ distribution, for THz wireless fiber extenders, was presented in~\cite{Boulogeorgos}. Expressions were deduced in~\cite{Bhardwaj}, corroborated by Monte-Carlo simulations, for the average SNR, ergodic capacity and ABER, considering a THz wireless system subject to $\alpha$-$\mu$ fading channel, pointing errors, and transceiver hardware impairments.
The authors in~\cite{AlBadarneh} present an analytical framework for capacity analysis of THz communication systems under various adaptive transmission techniques, in which exact expressions were derived by accounting for the effects of transceiver hardware impairments and modeling the channel using the H-function.
In the literature, note that most existing works rely on models that only partially capture the physical characteristics of THz channels.}

In this paper, a study of the $\alpha$-$\cal{F}$ fading distribution with beam misalignment, mobility \textcolor{black}{and hardware impairments} is performed, where new expressions for relevant statistics and metrics are derived. 
\textcolor{black}{The channel fading is modeled by the $\alpha$-$\cal{F}$ distribution~\cite{Badarneh}, that jointly considers the multipath fading and shadowing of a wireless fading channel. Additionally, it takes into consideration the non-linearity of the propagation medium. 
In our work, the $\alpha$-$\cal{F}$ model is considered since: 
(i) it is simple, characterized in terms of physical parameters, and generalist, encompassing other distributions as particular cases;
(ii) it characterizes small and large-scale fading, as well as the non-linearity of the communication channel and
(iii) it is supported by experimental results and adopted in many works under different scenarios, including THz communications~\cite{Almeida2023}.}
We also consider the recent beam misalignment model presented in~\cite{Dabiri}, suitable for millimeter wave to THz high-directional antenna arrays, taking into account the presence of unstable transmitter and receiver as a function of the antenna pattern.
Concerning mobility, the RWP is adopted, accounting for realistic distributions of users' positions, as presented in~\cite{Pereira}.
\textcolor{black}{To quantify the impact of hardware impairments on the system, the model presented in~\cite{Bjornson} is used, which has been thoroughly validated by means of theoretical and experimental approaches}.
Based on these characteristics, this work evaluates a THz wireless system that incorporates the combined effects of $\alpha$-$\cal{F}$ fading, beam misalignment, \textcolor{black}{hardware impairments} and large-scale factors such as mobility and distance-dependent path loss and atmospheric molecular attenuation.

The technical contributions of this paper are as follows: 
\begin{itemize}
    \item A study about the $\alpha$-$\cal{F}$ with beam misalignment, mobility \textcolor{black}{and hardware impairments} is performed, in which new closed-form expressions are derived for the PDF, CDF and higher-order moments of the instantaneous SNR;
    \item New expressions for OP, ASEP and channel capacity are derived; 
    \item \textcolor{black}{Asymptotic expressions are deduced in order to provide insights into the effect of the channel, mobility, hardware impairments and beam misalignment parameters on the high-SNR regime system performance. Furthermore, an upper bound on the capacity metric is obtained.}
\end{itemize}
\textcolor{black}{All expressions derived in this work are original and, to the best of our knowledge, this is the first work that jointly evaluates the effects of non-linearity of the propagation medium, shadowing, multipath, beam misalignment, mobility and hardware impairments within a unified framework. As a result, the analysis provides new insights into the behavior
of THz communication systems under multiple practical effects.}


The remainder of the paper is organized as follows. Section~\ref{system} describes the system, beam misalignment, mobility, \textcolor{black}{hardware impairments} and channel models adopted. In Section~\ref{modelo}, statistics are derived for the $\alpha$-${\cal{F}}$ composite fading distribution with beam misalignment, mobility \textcolor{black}{and hardware impairments}. Metrics and asymptotic metrics are presented in Sections~\ref{PerformanceAnalysis}. Section~\ref{results} shows the numerical results. Section~\ref{conclusao} brings the conclusions.

\color{black}
\subsection{Notation Remarks}

The complex Gaussian distribution is denoted as $\mathcal{CN}(\bar{\omega},\sigma^2)$, in which $\bar{\omega}$ and $\sigma^2$ are the corresponding mean and variance. The expected value operator is represented as $\mathbb{E}[\cdot]$. The Beta function~\cite[id. 06.18.02.0001.01]{Wolfram} is denoted by ${\rm B}(\cdot, \cdot)$. In turn, boldface letters represent vectors and $(\cdot)^{T}$ denotes the transpose operation. The Gamma function~\cite[id. 06.05.02.0001.01]{Wolfram} is written by~$\Gamma(\cdot)$. The Fox H-function~\cite[Eq. (1.2)]{Mathai} is represented by ${\rm H}_{p,q}^{m,n}\left[\cdot\right]$ and $\psi(z)$ is the digamma function~\cite{Weisstein}. $\delta(\cdot)$ and $\text{u}(\cdot)$ stands for the Dirac’s and Heaviside step functions, respectively.

\color{black}
\section{System, Beam  Misalignment, Channel, Mobility, and Hardware Impairments Models}\label{system}

\begin{figure}[!t]
    \begin{center}
    \includegraphics[width=0.85\linewidth]{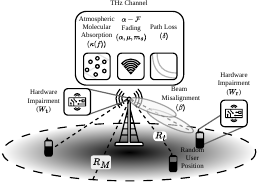}   
    \caption{\textcolor{black}{Illustration of the system model, considering the THz link affected by the atmospheric molecular absorption, $\alpha$-$\mathcal{F}$ fading, path losses, beam misalignment, hardware impairments, and random link distances.}}
\label{fig:system}
\end{center}
\end{figure}

The system under consideration is illustrated in Fig.~\ref{fig:system}, in which a downlink communication is established between a base station and a randomly positioned user, both equipped with high-directional antennas. \textcolor{black}{The link is impacted by the effects of $\alpha$-${\cal{F}}$ fading, path losses, beam misalignment, atmospheric molecular absorption and hardware impairments in both link sides. In addition, the effect of random distance between base station and user is also considered. In this configuration, the received signal is expressed as
\begin{equation}\label{eq:SinalRecebido}
Y_{r} = \sqrt{P_{\text{t}} G_{\text{t}} G_{\text{r}} \rho_{\text{l}} \rho_{\text{lm}} } H_\text{p} H_\text{f} (s + W_{\text{t}}) + W_{\text{r}} + W,
\end{equation}
in which $P_{\text{t}}$ denotes the transmitted power, $G_{\text{t}}$ and $G_{\text{r}}$ are the maximum antenna gains at transmitter and receiver, respectively; $s$ is the unit power transmitted symbol, and $W \sim \mathcal{CN}(0,\sigma_{W}^2)$ is the additive white Gaussian noise~(AWGN) with variance $\sigma_{W}^{2}$. 
}

In~(\ref{eq:SinalRecebido}), $H_\text{p}$ represents the pointing error, whose distribution of its instantaneous power, $Z=H_{\text{p}}^{2}$, is given by~\cite[Eq. (2)]{OsamahPoint}
\begin{equation}
f_Z(z) = -\frac{\beta^2}{4} \ln(z) z^{\frac{\beta}{2}-1}, \ 0 < z < 1,
\label{eq:fundZz}
\end{equation}
where $\beta = w_{B}^{2}/\sigma_{\theta}^{2}$, with $w_{B}$ denoting the antenna beamwidth and $\sigma_{\theta}^{2}$ the variance of antenna orientation fluctuations. According to this model, smaller values of $\beta$ represent more pronounced beam misalignments. In turn, the term $H_\text{f}$ in~(\ref{eq:SinalRecebido}) represents the composite fading envelope characterized by the $\alpha$-$\mathcal{F}$ distribution, whose PDF of the instantaneous power $X=\abs{H_{\text{f}}}^{2}$ can be written as~\cite{Badarneh}
\begin{align}
f_X(x) &= \frac{ \alpha }{ 2 \text{B}(\mu,m_s) } \left[  \frac{ (m_s-1) (\mathbb{E}[\abs{H_{\text{f}}}^{2}])^{\frac{\alpha}{2}} }{ \lambda^{\frac{\alpha}{2}} \mu } \right]^{m_s} \nonumber\\
    &\times x^{\frac{\alpha \mu}{2} - 1} \left[ x^{\frac{\alpha}{2}} + \frac{ (m_s-1) (\mathbb{E}[\abs{H_{\text{f}}}^{2}])^{\frac{\alpha}{2} } }{\lambda^{\frac{\alpha}{2}}\mu} \right]^{-(m_s+\mu)},
\label{eq:fundXx}    
\end{align}
in which $\mathbb{E}[\abs{H_{\text{f}}}^{2}]$ is its average power, $\alpha$ characterizes the non-linearity of the propagation medium, $\mu$ represents the number of multipath clusters and $m_s$ is the shadowing parameter. The $\alpha$-${\cal F}$ model can be reduced to characterize other types of fading, such as the Fisher-Snedecor, applying $\alpha=2$, and the $\alpha$-$\mu$, making $m_s\rightarrow \infty$.
Studies have experimentally validated the $\alpha$-$\mu$ distribution as a suitable model for characterizing small-scale fading in THz channels across diverse environments~\cite{Boulogeorgos,Papasotiriou2021b}. 

\textcolor{black}{The effect of mobility is analyzed assuming that the distance between transmitter and receiver is a continuous random variable~(RV) denoted as $R_{\text{t}}$. It is considered that $R_{\text{t}} = d_{0}D - d_{0}$, in which $D$ is a RV that takes a value $d$ in the interval $(1,1 + R_{M}/d_{0})$, with $d_{0}$ denoting a reference distance. From this formulation, $R_{\text{t}}$ is limited in the interval $[0,R_{M}]$, with $R_{M}$ representing the maximum separation distance between transmitter and receiver. According to the RWP model, the RV $D$ has a PDF given by~\cite[Eq. (2)]{Pereira}
\begin{equation}
f_{D}(d) = \sum_{i=1}^{l}B_i\left(\frac{d_0}{R_M}\right)^{\beta_i+1}(d-1)^{\beta_i},
\label{eq:fDundd}
\end{equation}
in which the parameters $l$, $B_i$ and $\beta_i$ depend on the number of dimensions considered in the topology. For instance, the one-dimensional~(1D) case is parameterized with $l = 2$, $B_i = [6, -6]$, and $\beta_i = [1, 2]$. The two-dimensional~(2D) topology is modeled by $l=3$, $B_i = (1/73)[324, -420, 96]$
and $\beta_i = [1, 3, 5]$; and the three-dimensional~(3D) topology is expressed by $l = 3$, $B_i = (1/72)[735, -1190, 455]$ and $\beta_{i} = [2,4,6]$.}

The term $\rho_{\text{l}}$ represents the \textcolor{black}{distance-dependent} path losses and is calculated by
\begin{equation}
    \rho_{\text{l}} =  \left(\frac{\lambda_{\text{0}}}{4\pi d_{0}}\right)^2 \left(\frac{R_{t}}{d_{0}}\right)^{-\delta},
\end{equation}
where $\lambda_{\text{0}}$ is the wavelength and $\delta$ is the path loss exponent. \textcolor{black}{In turn}, the \textcolor{black}{distance-dependent} atmospheric absorption factor, denoted as $\rho_{\text{lm}}$, is given by 
\begin{equation}
  \rho_{\text{lm}} = \exp[-\kappa(f)R_{t}],
\end{equation}
where $\kappa(f)$ is the frequency-dependent specific attenuation coefficient, which depend on the local temperature, pressure and water vapor density. \textcolor{black}{This coefficient can be calculated based on the ITU-R Recommendation P.676}~\cite{ITU}\footnote{Recommendation ITU-R P.676 presents a methodology for calculating the specific attenuation coefficient expressed in dB/km. To adapt this coefficient to the formulation adopted in this work, we consider that $\kappa(f) = \ln(10) 10^{-5} \kappa_{0}(f)$, in which $\kappa_{0}(f)$ is the ITU-R based coefficient expressed in dB/km.}.

\textcolor{black}{Finally, the effects of hardware impairments on the received signal, which take into account deviation in the transmitted symbol and reduction in SNR experienced by the receiver; are statistically modeled using the RVs $W_{\text{t}} \sim \mathcal{CN}(0,k_{\text{t}}^2)$ and $W_{\text{r}}\sim \mathcal{CN}(0,k_{\text{r}}^2 P_{\text{t}} G_{\text{t}} G_{\text{r}} \rho_{\text{l}} \rho_{\text{lm}} {H_\text{p}}^{2} \abs{H_\text{f}}^2 )$, where $k_{\text{t}}$ and $k_{\text{r}}$ represent the degree of hardware impairments in the transmitter and receiver, respectively~\cite{Bjornson}.} 




\color{black}
Without loss of generality, it is assumed that the joint pointing error-fading power is normalized to $\mathbb{E}[H_{p}^{2} \abs{H_{f}}^{2}]=1$. Therefore, the average received symbol power is given by 
\begin{equation}
    \Omega = \mathbb{E}\left[ \abs*{ \sqrt{P_{\text{t}} G_{\text{t}} G_{\text{r}} \rho_{\text{l}} \rho_{\text{lm}} } H_\text{p} H_\text{f} s }^2 \right] = P_{\text{t}} G_{\text{t}} G_{\text{r}} \rho_{\text{l}} \rho_{\text{lm}}.
\end{equation}
Considering the dependence of the terms $\rho_{\text{l}}$ and $\rho_{\text{lm}}$ on the distance and that $R_{\text{t}}=d_{0} D - d_{0}$, the average power of the received symbol conditioned on the distance $d$ can be expressed as
\begin{equation}
    \label{eqomegad}
    \Omega(d) = \Omega_{0} \left( d - 1 \right)^{-\delta}  \exp[-\kappa(f) d_{0}d  ],
\end{equation}
where 
\begin{equation}
    \Omega_0=P_{\text{t}}G_{\text{t}} G_{\text{r}}\left(\frac{\lambda_{\text{0}}}{4\pi d_{0}}\right)^2\exp[\kappa(f)d_0].   
\end{equation}
In turn, the power of the noise plus the effects of hardware impairments conditioned to a distance $d$, denoted as $\Theta(d)$; is calculated as
\begin{equation}
    \label{thetad}
    \begin{aligned}
    \Theta(d) &= \mathbb{E}\left[ \abs*{ \sqrt{P_{\text{t}} G_{\text{t}} G_{\text{r}} \rho_{\text{l}} \rho_{\text{lm}} } H_\text{p} H_\text{f} W_{t} + W_{\text{r}} + W }^2 \right] \\
    &= \Omega(d) (k_{\text{t}}^2 + k_{\text{r}}^2) + \sigma_{W}^2.
    \end{aligned}
\end{equation}


Considering~\eqref{eqomegad} and~\eqref{thetad}, the instantaneous SNR, conditioned on distance $d$, can be expressed as 
\begin{equation}
    \label{eq:snr}
    \begin{aligned}
    \Gamma(d) &\triangleq \frac{\Omega(d) H_{\text{p}}^2 \abs{ H_{\text{f}}}^2}{\Omega(d) H_{\text{p}}^2 \abs{ H_{\text{f}}}^2 (k_{\text{t}}^2 + k_{\text{r}}^2) + \sigma_{W}^2} \\
    &=\frac{\Omega_{0} \left( d - 1 \right)^{-\delta}  \exp[-\kappa(f) d_{0}d] H_{\text{p}}^2 \abs{ H_{\text{f}}}^2 }{\Omega_{0} \left( d - 1 \right)^{-\delta}  \exp[-\kappa(f) d_{0}d] H_{\text{p}}^2 \abs{ H_{\text{f}}}^2 (k_{\text{t}}^2 + k_{\text{r}}^2) + \sigma_{W}^2} \\
    &= \frac{\gamma_{0} \left( d - 1 \right)^{-\delta}  \exp[-\kappa(f) d_{0}d] H_{\text{p}}^2 \abs{ H_{\text{f}}}^2 }{\gamma_{0} \left( d - 1 \right)^{-\delta}  \exp[-\kappa(f) d_{0}d] H_{\text{p}}^2 \abs{ H_{\text{f}}}^2 k^2 + 1},
    \end{aligned}
\end{equation}
where $k^2 = k_{\text{t}}^2 + k_{\text{r}}^2$ represents the joint level of hardware impairments and
\begin{equation}
    \label{eq:snr_0}
    \gamma_{0} = \frac{\Omega_{0}}{\sigma_{W}^2} = \bar{\gamma} \left(\frac{\lambda_{\text{0}}}{4\pi d_{0}}\right)^2\exp[\kappa(f)d_0],
\end{equation}
with $\bar{\gamma} = P_{\text{t}}G_{\text{t}} G_{\text{r}} / \sigma_{W}^2$. Analyzing~\eqref{eq:snr} and~\eqref{eq:snr_0}, we have that $\Gamma(d)=0$ in the case where $\bar{\gamma} = 0$ and that $\Gamma(d) \rightarrow 1/k^2$ when $\bar{\gamma} \rightarrow \infty$. Therefore, the SNR is limited by the hardware impairments level, such that $0\leq \Gamma(d) \leq 1/k^2$.

\color{black}

\color{black}
\section{The $\alpha$-${\cal{F}}$ Distribution under Beam Misalignment, Mobility, and Hardware Impairments}
\label{modelo}

\begin{lemma} \label{eq:lema1}
(PDF and CDF of the SNR $\Gamma$) -- For $\alpha$, $\mu$, $k$, $\beta$, $\lambda$, $m_s$, $\delta$, $\gamma_{0}$ $\in$ $\mathbb{R}^{+}$, $m_s>\max( 2/\alpha,1)$, $\beta>2$, $\delta \in [2,6]$, with $N$ being the number of weights $w_{k}$ of 
a Gauss-Legendre quadrature\footnote{\textcolor{black}{We employ the Gauss–Legendre quadrature in this work, which approximates definite integrals as weighted sums of function evaluations at specific nodes. This technique is widely adopted in similar contexts and provides deterministic and reproducible results for a fixed number of quadrature points, with implementations available across multiple programming environments. Moreover, Gauss–Legendre quadrature yields highly accurate numerical approximations by using a reduced number of roots and weights.}} and $x_{k}$ the $k$-th root of a Legendre polynomial of order $N$, the PDF and CDF of the SNR for the $\alpha$-$\mathcal{F}$ 
fading model with beam misalignment, mobility and hardware impairments are expressed respectively by~(\ref{eq:mobilityPDF}) and~(\ref{eq:mobilityCDF}), given at the top of the next page, in which 
\begin{equation}
    \mathbf{v}_1^T(i)=\left[\frac{R_M}{2d_0},\,\,\beta_i+\frac{\alpha\mu\delta}{2},\,\,\frac{\alpha}{2}\kappa(f)\mu d_0\right],
\end{equation}
\begin{equation}
    \mathbf{v}_2^T=\left[\frac{R_M}{2d_0},\,\,\frac{\alpha \delta}{2},\,\,\frac{\alpha}{2}\kappa(f)d_0\right],
\end{equation}
\begin{equation}
\psi(a,b,x)=\frac{a(x+1)}{2\left(1-\frac{ab}{2}(x+1)\right)},    
\end{equation}
\begin{align}
    \Phi(x_k;\mathbf{a}^T)&=[a_1(x_k+1)]^{a_2} \exp\{a_3[a_1(x_k+1)+1]\},
\end{align}
where $\mathbf{a}^{T}=
[a_{1},\,\,a_{2},\,\,a_{3}]$ and the constants $c_1$ and $c_2$ are given by~(\ref{eq:Constantec1}) and~(\ref{eq:Constantec2}), respectively.
\end{lemma}


\begin{proof}
See Appendix \ref{FirstAppendix}.
\end{proof}

\begin{figure*}
\textcolor{black}{
\begin{align}\label{eq:mobilityPDF}
f_{\Gamma}(\gamma) &= \frac{2R_M c_1 c_2^{-(\mu+m_s)}}{d_0\Gamma(\mu+m_s)}
\frac{(\psi(2,k^2,\gamma-1))^{\frac{\alpha\mu}{2}+1}}{\gamma_{0}^{\frac{\alpha\mu}{2}}\gamma^{2}}
\sum_{i=1}^{l}\sum_{k=1}^{N} B_i \left( \frac{d_0}{R_M} \right)^{\beta_i+1} w_k \Phi(x_k;\mathbf{v}_1^T(i))\nonumber \\
    &\times{\rm H}_{3,3}^{3,1}\left[ \frac{\Phi(x_k;\mathbf{v}_2^T)}{c_2\gamma_{0}^{\frac{\alpha}{2}}}
    \left(\psi(2,k^2,\gamma-1)\right)^{\frac{\alpha}{2}} \middle\vert 
     \begin{array}{c}
        (1-\mu-m_s,1),(1+\beta -\alpha \mu, \alpha), (1+\beta-\alpha \mu, \alpha)  \\
        (0,1), (\beta-\alpha \mu, \alpha), (\beta-\alpha \mu, \alpha)
     \end{array}\right], \\
     &\text{if $0\leq\gamma <1/k^2$ and 0 otherwise.} \nonumber
\end{align}}    
\hrulefill
\textcolor{black}{
\begin{align}\label{eq:mobilityCDF}
 F_{\Gamma}(\gamma) &= \frac{R_M c_1 c_2^{-(\mu+m_s)}}{d_0\Gamma(\mu+m_s)}\frac{\gamma}
 {\gamma_{0}^{\frac{\alpha\mu}{2}}}\sum_{i=1}^{l}\sum_{k_{1}=1}^{N}\sum_{k_{2}=1}^{N} B_i \left( \frac{d_0}{R_M} \right)^{\beta_i+1} w_{k_{1}}w_{k_{2}} \Phi(x_{k_{1}};\mathbf{v}_1^T(i))
 \frac{(\psi(\gamma,k^{2},x_{k_{2}}))^{\frac{\alpha\mu}{2}+1}}{(\frac{\gamma}{2}(x_{k_{2}}+1))^{2}} \nonumber \\
    &\times {\rm H}_{3,3}^{3,1}\left[ \frac{\Phi(x_{k_{1}};\mathbf{v}_2^T)}{c_2} 
    \left(\frac{\psi(\gamma,k^{2},x_{k_{2}})}{\gamma_{0}}\right)^{\frac{\alpha}{2}} 
    \middle\vert 
     \begin{array}{c}
        (1-\mu-m_s,1), (1+\beta -\alpha \mu, \alpha), (1+\beta-\alpha \mu, \alpha)  \\
        (0,1), (\beta-\alpha \mu, \alpha), (\beta-\alpha \mu, \alpha)
     \end{array}\right], \\
     &\text{if $0\leq\gamma <1/k^2$ and 1 otherwise.} \nonumber
\end{align}
}
\hrulefill
\end{figure*}

\begin{lemma} \label{eq:lema2}
(Higher-Order Moments of the SNR $\Gamma$) -- For the instantaneous SNR $\Gamma$, characterized by the parameters and
distribution stated in {\bf Proposition \ref{eq:lema1}}, the $n$-th moment is expressed by (\ref{eq:rawmoment}).
\end{lemma}

\begin{proof}
See Appendix \ref{SecondAppendix}.
\end{proof}

\begin{figure*}
\textcolor{black}{
\begin{align}\label{eq:rawmoment}
\mathbb{E}[\Gamma^{n}]&=\frac{2R_{M}c_{1}c_{2}^{-(\mu+m_{s})}}{\Gamma(n)k^{2n+\alpha\mu}d_{0}\Gamma(\mu+m_{s})
\gamma_{0}^{\frac{\alpha\mu}{2}}}\sum_{i=1}^{l}\sum_{k=1}^{N}B_{i}\left(\frac{d_{0}}{R_{M}}\right)^{\beta_{i}+1}
w_{k}\Phi(x_{k};\mathbf{v}_{1}^{T}(i))\nonumber \\
&\times \text{H}_{4,4}^{4,2}\left[\left.\frac{\Phi(x_{k},\mathbf{v}_{2}^{T})}
{c_{2}(k^{2}\gamma_{0})^{\frac{\alpha}{2}}}\right|\begin{array}{c}
(1-m_{s}-\mu,1), (1-n-\frac{\alpha\mu}{2},\frac{\alpha}{2}), (1+\beta-\alpha\mu,\alpha), (1+\beta-\alpha\mu,\alpha) \\
(0,1), (\beta-\alpha\mu,\alpha), (\beta-\alpha\mu,\alpha), (-\frac{\alpha\mu}{2},\frac{\alpha}{2})
\end{array}\right]
\end{align}
}
\hrulefill
\end{figure*}

\color{black}
\section{Performance Analysis}\label{PerformanceAnalysis}

In this section, the analytical solutions for OP, ASEP and average capacity of the considered system are developed.

\subsection{Outage Probability}

The OP, denoted by $P_{\rm{out}}$, can be obtained directly from (\ref{eq:mobilityCDF}) as $P_{\rm{out}} = F_{\Gamma}(\gamma_{\rm{th}})$, in which $\gamma_{\textrm{th}}$ is the SNR threshold. \textcolor{black}{It is important to note that hardware impairments impose an upper bound $\Gamma \leq 1/k^2$. Consequently, when $\gamma_{\text{th}} > 1/k^2$, the OP is necessarily equal to $P_{\text{out}} = 1$.}

\textcolor{black}{The asymptotic behavior of the OP can be derived by expanding the Fox H-function of $F_{\Gamma}(\gamma)$ as a sum of residuals of the Mellin-Barnes integrand, evaluated at the principal poles of the gamma functions that are to the left of the integration path~\cite{HTransforms}. Among these functions, there exists one squared gamma function exhibiting poles $\zeta_{l2}$ of multiplicity 2, as well as one gamma function with simple poles $\zeta_{l1}$, $l=0,1,\cdots$. For the squared gamma function, the residue at its principal pole $\zeta_{02}=-(\beta-\alpha\mu)/\alpha$ of multiplicity 2 is given by
\begin{equation}
\mathrm{Res}(\zeta_{02})=\lim_{\zeta\rightarrow \zeta_{02}}\frac{d}{d\zeta}
\left[(\zeta-\zeta_{02})^{2}\Gamma^{2}(\beta-\alpha\mu+\alpha \zeta)\mathcal{H}_{2}(\zeta)\right],
\label{eqres02}
\end{equation}
in which 
\begin{equation}
\frac{\mathcal{H}_{2}(\zeta)}{\Lambda_{2}(\zeta)}=
\exp\left[-\zeta \ln\left(\frac{\phi(x_{k_{1}},\mathrm{\textbf{v}}_{2}^{T})}{c_{2}}
\left(\frac{\psi(\gamma,k^{2},x_{k_{2}})}{\gamma_{0}}\right)^{\alpha/2}\right)\right],
\end{equation}
and 
\begin{equation}
\Lambda_{2}(\zeta)=\frac{\Gamma(\zeta)\Gamma(\mu+m_{s}-\zeta)}
    {\Gamma^{2}(1+\beta-\alpha\mu+\alpha \zeta)}.
\end{equation}
}

\textcolor{black}{For $\Gamma(\zeta)$, the residue at its principal pole $\zeta_{01}=0$ is given by
\begin{equation}
\mathrm{Res}(\zeta_{01})=\lim_{\zeta\rightarrow \zeta_{01}}(\zeta-\zeta_{01})\Gamma(\zeta)\mathcal{H}_{1}(\zeta),
\label{eqres01}
\end{equation}
in which
\begin{equation}
\mathcal{H}_{1}(\zeta)=\Lambda_{1}(\zeta)\left(\frac{\Phi(x_{k},\mathrm{\textbf{v}}_{2}^{T})}{c_{2}}
\left(\frac{\psi(\gamma,k^{2},x_{k_{2}})}{\gamma_{0}}\right)^{\frac{\alpha}{2}}\right)^{-\zeta}
\end{equation}
and 
\begin{equation}
\Lambda_{1}(\zeta)=\frac{\Gamma^{2}(\beta-\alpha\mu+\alpha\zeta)\Gamma(\mu+m_{s}-\zeta)}
    {\Gamma^{2}(1+\beta-\alpha\mu+\alpha \zeta)}.
\end{equation}
}

\textcolor{black}{Then, by calculating the derivative required for the residue at the principal pole of multiplicity 2 and selecting the terms with the highest growth, the asymptotic expression of the CDF $F_{\Gamma}(\gamma)$ in terms of the dominant residue for $\beta>\alpha\mu$ is obtained as
\begin{align}
  F_{\Gamma}^{\infty}(\gamma) &\simeq \frac{2c_1 c_2^{-(\mu+m_{s})}}{(\beta-\alpha \mu)^2}
  \left(\frac{\gamma}{2\gamma_{0}}\right)^{\frac{\alpha\mu}{2}}
\sum_{i=1}^{l}\sum_{k_{1}=1}^{N}\sum_{k_{2}=1}^{N} B_i \left( \frac{d_0}{R_M} \right)^{\beta_i} \nonumber \\
    &\times w_{k_{1}}w_{k_{2}} \Phi(x_{k_{1}};\mathbf{v}_1^T(i))\frac{(1+x_{k_{2}})^{\frac{\alpha\mu}{2}-1}}
    {\left(1-k^{2}\frac{\gamma}{2}(1+x_{k_{2}})\right)^{\frac{\alpha\mu}{2}+1}}
\end{align}
and for $\beta\leq \alpha\mu$, the asymptotic expression of $F_{\Gamma}(\gamma)$ in terms of the dominant residue is
\begin{align}
&F_{\Gamma}^{\infty}(\gamma) \simeq \frac{2c_1 c_2^{-(m_{s}+\mu)}}{\alpha^{2}\Gamma(\mu+m_{s})}
\Lambda_{2}\left(\mu-\frac{\beta}{\alpha}\right)\left(\frac{\gamma}{2\gamma_{0}}\right)^{\frac{\beta}{2}}\nonumber\\
&\times\sum_{i=1}^{l}\sum_{k_{1}=1}^{N}\sum_{k_{2}=1}^{N} B_i \left( \frac{d_0}{R_M} \right)^{\beta_i} w_{k_{1}}w_{k_{2}} \Phi(x_{k_{1}};\mathbf{v}_1^T(i))
\nonumber\\
&\times\left( \frac{\Phi(x_{k_{1}};\mathbf{v}_2^T)}{c_2}\right)^{\frac{\beta}{\alpha}-\mu}
\frac{(1+x_{k_{2}})^{\frac{\beta}{2}-1}}{\left(1-\frac{k^{2}}{2}\gamma(1+x_{k_{2}})\right)^{\frac{\beta}{2}+1}}\\
&\times\left\{\frac{\alpha}{2}\ln\left(\frac{2\gamma_{0}}{\gamma}\right)-\right.
\ln\left(\frac{\Phi(x_{k_{2}};\mathbf{v}_{2}^{T})}{c_{2}}\right)\nonumber \\
&-\left.\frac{\alpha}{2}\ln\left(\frac{1+x_{k_{2}}}{\left(1-\frac{k^{2}}{2}\gamma(1+x_{k_{2}})\right)}\right)\right\}.
\end{align}
Therefore, the asymptotic behavior of the OP is expressed as $P_{\textrm{out}}^{\infty}=F_{\Gamma}^{\infty}(\gamma_{\mathrm{th}})$, with diversity gain $G_d = \frac{\alpha\mu}{2}$ for $\beta>\alpha\mu$; and $G_d = \frac{\beta}{2}$ for $\beta\leq \alpha\mu$.}

\subsection{Average SEP for Binary Constellations}

The ASEP, denoted by $P_{s}$, can be calculated by means of~\cite[Eq. (15)]{OsamahPoint}
\begin{equation}\label{eq:FormulaSEP}
P_{s}=\frac{a\sqrt{b}}{2\sqrt{2\pi}}\int_{0}^{\infty}\gamma^{-\frac{1}{2}}F_{\Gamma}(\gamma)e^{-\frac{b}{2}\gamma}\text{d}\gamma,
\end{equation}
\textcolor{black}{where the parameters $a$ and $b$ depend on the adopted modulation scheme. For instance, $(a,b)=(1,2)$ corresponds to binary phase-shift keying (BPSK). The integral expression in~\eqref{eq:FormulaSEP} is exact for binary constellations; however, it also provides an accurate high-SNR approximation for non-binary constellations, using the corresponding values of $a$ and $b$.}

\textcolor{black}{
To evaluate~\eqref{eq:FormulaSEP}, the CDF $F_{\Gamma}(\gamma)$ [\textit{cf.}~(\ref{eq:mobilityCDF})] is first expressed in terms of its Mellin–Barnes integral representation, leading to~\eqref{eq:asepmelin}. The resulting definite integral with respect to $\gamma$ is then solved using Gauss–Legendre quadrature. By reformulating the improper integral in terms of the complementary error function and subsequently expressing the result via the Fox H-function~\cite[Eq. (1.2)]{Mathai}, $P_{s}$ can be obtained in the form given in (\ref{eq:mobilityASEP}), where
\begin{equation}
w_{k_{2}}^{\prime}=w_{k_{2}}(1+x_{k_{2}})^{-2},   
\end{equation}
\begin{equation}
w_{k_{3}}^{\prime}=w_{k_{3}}(1+x_{k_{3}})^{-(\alpha\mu/2+1)},    
\end{equation}
\begin{equation}
\varphi(x_{k_{2}},x_{k_{3}})=\frac{\frac{1}{2}(1+x_{k_{2}})(1+x_{k_{3}})}{\left(1-\frac{1}{4}(1+x_{k_{2}})(1+x_{k_{3}})\right)},    
\end{equation}
\begin{equation}
\label{eq:c3}
c_{3}=\frac{a}{2}\mathrm{erfc}\left(\sqrt{\frac{b}{2k^{2}}}\right),    
\end{equation}
and
\begin{equation}
\mathbf{v}_{3}^{T}=\left[\frac{1}{2k^{2}},\frac{\alpha\mu}{2}-\frac{1}{2},-\frac{b}{2}\right],    
\end{equation}
with $w_{k_{u}}$, with $u\in\{1,2,3\}$, being the $k_{u}$-th weight of Gauss-Legendre quadrature and $x_{k_{u}}$ being the $k_{u}$-th root of the Legendre polynomial of degree $N$.
}

\begin{figure*}
\textcolor{black}{
\begin{align}\label{eq:asepmelin}
P_{s}&= \frac{a\sqrt{b} R_M c_1 c_2^{-(\mu+m_s)}}{2\sqrt{2\pi}d_0\Gamma(\mu+m_s)\gamma_{0}^{\frac{\alpha\mu}{2}}}
\sum_{i=1}^{l}\sum_{k_{1}=1}^{N}\sum_{k_{2}=1}^{N}B_{i}\left(\frac{d_{0}}{R_{M}}\right)^{\beta_{i}+1}w_{k_{1}}w_{k_{2}}
\Phi(x_{k_{1}},\mathbf{v}_{1}^{T}(i))\left(\frac{1}{2}(x_{k_{2}}+1)\right)^{\frac{\alpha\mu}{2}-1}\nonumber\\
&\times\frac{1}{j2\pi}\int_{L}\frac{\Gamma(\zeta)\Gamma^{2}(\beta-\alpha\mu+\alpha\zeta)\Gamma(1-(1-\mu-m_s)-\zeta)}
{\Gamma^{2}(1+\beta-\alpha\mu+\alpha\zeta)}\left(\frac{\Phi(x_{k_{1}};\mathbf{v}_2^T)}{c_2 \gamma_{0}^{\frac{\alpha}{2}}}
\left(\frac{1}{2}(1+x_{k_{2}})\right)^{\frac{\alpha}{2}}\right)^{-\zeta}\nonumber\\ 
&\times\int_{0}^{\frac{1}{k^{2}}}\frac{\gamma^{\frac{\alpha\mu}{2}-\frac{\alpha \zeta}{2}-\frac{1}{2}}}
{\left(1-\frac{k^{2}}{2}(1+x_{k_{2}})\gamma\right)^{\frac{\alpha\mu}{2}-\frac{\alpha \zeta}{2}+1}}e^{-\frac{b}{2}\gamma}
\textrm{d}\gamma\textrm{d}\zeta+\frac{a\sqrt{b}}{2\sqrt{2\pi}}\int_{\frac{1}{k^{2}}}^{\infty}\gamma^{-1/2}e^{-\frac{b}{2}\gamma}\mathrm{d}\gamma.
\end{align}
}
\hrulefill
\end{figure*}

\begin{figure*}
\textcolor{black}{
\begin{align}\label{eq:mobilityASEP}
    P_s &= \frac{a\sqrt{b} R_M c_1 c_2^{-(\mu+m_s)}e^{\frac{b}{2}}}{\sqrt{2\pi}d_0\Gamma(\mu+m_s)k^{2}\gamma_{0}^{\frac{\alpha\mu}{2}}}
   \sum_{i=1}^{l}\sum_{k_{1}=1}^{N}\sum_{k_{2}=1}^{N}\sum_{k_{3}=1}^{N} B_i \left( \frac{d_0}{R_M} \right)^{\beta_i+1} 
   w_{k_{1}}w_{k_{2}}^{\prime}w_{k_{3}}^{\prime}\frac{\Phi(x_{k_{1}};\mathbf{v}_1^T(i))\Phi(x_{k_{3}};\mathbf{v}_3^T)}
   {(\varphi(x_{k_{2}},x_{k_{3}}))^{-(\frac{\alpha\mu}{2}+1)}}\nonumber \\
    &\times {\rm H}_{3,3}^{3,1}\left[ \frac{\Phi(x_{k_{1}};\mathbf{v}_2^T)}{c_2 \left(2\gamma_0 k^{2}\right)^{\frac{\alpha}{2}}}
    (\varphi(x_{k_{2}},x_{k_{3}}))^{\frac{\alpha}{2}} \middle\vert 
     \begin{array}{c}
        (1-\mu-m_s,1), (1+\beta -\alpha \mu, \alpha), (1+\beta-\alpha \mu, \alpha) \\
        (0,1), (\beta-\alpha \mu, \alpha), (\beta-\alpha \mu, \alpha)
     \end{array}\right]+c_{3}.
\end{align}   
}
\hrulefill
\end{figure*}

{\color{black}
The high SNR behavior of ASEP can be derived knowing that $\lim_{\gamma_{0}\rightarrow \infty}\Gamma \rightarrow 1/k^{2}$, which indicates that~\cite{Boulogeorgos}
\begin{equation}
\label{eq:pdf_hsnr}
\lim_{\gamma_{0}\rightarrow \infty}f_{\Gamma}(\gamma)\rightarrow
\delta\left(\gamma-\frac{1}{k^{2}}\right),
\end{equation}
and
\begin{equation}
    \label{eq:cdf_hsnr}
    \lim_{\gamma_{0}\rightarrow \infty} F_{\Gamma}(\gamma) \rightarrow \text{u}(\gamma - 1/k^2).
\end{equation}
Applying~\eqref{eq:cdf_hsnr} to~\eqref{eq:FormulaSEP}, it can be demonstrated that in a high SNR regime, ASEP converges to an irreducible level equal to
\begin{equation}
    \begin{aligned}
    \label{eq:asep_hsnrlim}
    P_{s}^{\infty} = \lim_{\gamma_{0}\rightarrow \infty} P_{s} &= \frac{a\sqrt{b}}{2\sqrt{2\pi}}\int_{\frac{1}{k^2}}^{\infty}\gamma^{-\frac{1}{2}}e^{-\frac{b}{2}\gamma}\text{d}\gamma = c_3,
    \end{aligned}
\end{equation}
in which $c_{3}$ is given by (\ref{eq:c3}).
}


\subsection{Average Channel Capacity}

{\color{black}

The average channel capacity is given by
\begin{equation}\label{eq:CapacidadeFormula}
C=\frac{1}{\ln(2)}\int_{0}^{\infty}\ln(1+\gamma)f_{\Gamma}(\gamma)\mathrm{d}\gamma.
\end{equation}
To evaluate this integral, we substitute~\eqref{eq:mobilityPDF} into~\eqref{eq:CapacidadeFormula} and rewrite it in the form of~\eqref{eq:mobilityEC1}. The integration interval is then transformed from $[0,1/k^{2}]$ to $[-1,1]$ using~\cite[Eq. (3.021.1)]{Gradsteyn2007}, enabling the application of Gauss–Legendre quadrature. As a result, $C$ can be expressed as in~\eqref{eq:mobilityEC2}. An upper bound on the average channel capacity follows from Jensen’s inequality:
\begin{equation}
C=\frac{1}{\ln(2)}\mathbb{E}[\ln(1+\Gamma)]\leq\frac{1}{\ln(2)}\ln\bigl(1+\mathbb{E}[\Gamma]\bigr),
\end{equation}
where $\mathbb{E}[\Gamma]$ is obtained from~\eqref{eq:rawmoment} by setting $n=1$.

The asymptotic behavior of the capacity is determined by substituting~\eqref{eq:pdf_hsnr} into~\eqref{eq:CapacidadeFormula}, yielding
\begin{equation}
    \begin{aligned}
    C^{\infty} &= \lim_{\gamma_{0}\rightarrow \infty} C = \int_{0}^{\infty}\log_{2}(1+\gamma)\delta\left(\gamma-\frac{1}{k^{2}}\right)\mathrm{d}\gamma\\ 
    &= \log_{2}\left(1+\frac{1}{k^{2}}\right),
    \end{aligned}
\end{equation}
which shows that the capacity is fundamentally limited by the level of transceiver hardware impairments.


}
\begin{figure*}
\textcolor{black}{
\begin{align}\label{eq:mobilityEC1}
C &=\frac{2R_Mc_1 c_2^{-(\mu+m_s)}}{\ln(2)d_0\Gamma(\mu+m_s)\gamma_0^{\frac{\alpha \mu}{2}}}
\sum_{i=1}^{l}\sum_{k_{1}=1}^{N} B_i \left( \frac{d_0}{R_M} \right)^{\beta_i+1}w_{k_{1}}\Phi(x_{k_{1}};\mathbf{v}_1^T(i)) 
\int_{0}^{\frac{1}{k^{2}}}\frac{\gamma^{\frac{\alpha\mu}{2}-1}\ln(1+\gamma)}{(1-k^{2}\gamma)^{\frac{\alpha\mu}{2}+1}}\nonumber\\
&\times {\rm H}_{3,3}^{3,1}\left[\frac{\Phi(x_{k_{1}},\mathbf{v}_{2}^{T})}{c_{2}\gamma_{0}^{\frac{\alpha}{2}}}\left(\frac{\gamma}{1-k^{2}\gamma}\right)^{\frac{\alpha}{2}} \middle\vert 
     \begin{array}{c}
        (1-m_s-\mu,1), (1+\beta-\alpha\mu,\alpha), (1+\beta-\alpha\mu,\alpha)  \\
        (0,1), (\beta-\alpha\mu,\alpha), (\beta-\alpha\mu,\alpha)
\end{array}\right]\text{d}\gamma, 
\end{align}    
}
\hrulefill
\end{figure*}

\begin{figure*}
\textcolor{black}{
\begin{align}\label{eq:mobilityEC2}
C&=\frac{R_M c_1 c_2^{-(\mu+m_s)}}{\ln(2)d_0\Gamma(\mu+m_s)(k^2\gamma_{0})^{\frac{\alpha \mu}{2}}}
\sum_{i=1}^{l}\sum_{k_{1}=1}^{N}\sum_{k_{2}=1}^{N}B_i\left(\frac{d_0}{R_M} \right)^{\beta_i+1} 
w_{k_{1}}w_{k_{2}}\nonumber\\
&\times\left(\frac{1}{2}(1+x_{k_{2}})\right)^{-2}\ln\left(1+\frac{1}{2k^{2}}(1+x_{k_{2}})\right)
\Phi(x_{k_{1}};\mathbf{v}_1^T(i))\left(\psi(1,1,x_{k_{2}})\right)^{\frac{\alpha\mu}{2}+1}\nonumber\\ 
 &\times{\rm H}_{3,3}^{3,1}\left[\frac{\Phi(x_k;\mathbf{v}_2^T)}{c_2(k^{2}\gamma_{0})^{\frac{\alpha}{2}}}
 (\psi(1,1,x_{k_{2}}))^{\frac{\alpha}{2}}\middle\vert 
     \begin{array}{c}
        (1-\mu-m_s,1), (1+\beta -\alpha \mu, \alpha), (1+\beta-\alpha \mu, \alpha) \\
        (0,1), (\beta-\alpha \mu, \alpha), (\beta-\alpha \mu, \alpha)
     \end{array}\right]
\end{align}    
}
\hrulefill
\end{figure*}
\sethlcolor{green}

\section{Numerical Results}
\label{results}

\textcolor{black}{
This section presents analytical results for the performance metrics considered in this work. The evaluated metrics are affected by $\alpha$–$\mathcal{F}$ fading, hardware impairments, beam misalignment, atmospheric molecular absorption, and path loss, while also accounting for the impact of random link distance. The derived analytical expressions are validated through Monte Carlo simulations that replicate the system model described in Section~\ref{system}. It is important to note that the analytical expressions for the OP, ASEP, and ergodic capacity are expressed as weighted sums of Fox’s $\text{H}$-functions, which may impose a significant computational burden. However, from~\eqref{eq:mobilityPDF},~\eqref{eq:mobilityCDF},~\eqref{eq:mobilityASEP}, and~\eqref{eq:mobilityEC2}, it can be observed that all H-functions share the same parameter structure and identical parameter sets $(\alpha, \mu, m_{s}, \beta)$. This property enables precomputation of the function over a sufficiently dense range of its argument, allowing subsequent evaluations to be performed via lookup (e.g., nearest-neighbor search) in the precomputed table. As a result, the number of quadrature roots, responsible for most of the summation terms; can be flexibly adjusted without incurring the high computational cost associated with repeated evaluations of the Fox H-function.}

\textcolor{black}{
In the path loss parameterization, the reference distance is set to $d_{0}=1$~m, and the path loss exponent $\delta$ is assumed to lie within the interval $[2,4]$, which is the typical range for the evaluated scenario, as reported in~\cite{Rappaport2019}. To determine the specific attenuation due to atmospheric molecular absorption, standard atmospheric conditions are considered, namely a temperature of 290~K, water vapor density of 7.5~g/m$^3$, and pressure of 1013.25~hPa~\cite{ITU}. A high-gain directional link is analyzed, with antenna gains $G_{\text{t}} = G_{\text{r}} = 55$~dBi~\cite{Boulogeorgos}. Since the performance metrics depend on the reference SNR $\bar{\gamma} = P_{\text{t}} G_{\text{t}} G_{\text{r}} / \sigma_{W}^{2}$, the results are presented as functions of the normalized transmit power $P_{\text{t}}/\sigma_{W}^{2}$. This representation enables flexible evaluation across different system configurations that yield the same ratio.}

\begin{figure}
\centering
\includegraphics[width=0.7\linewidth]{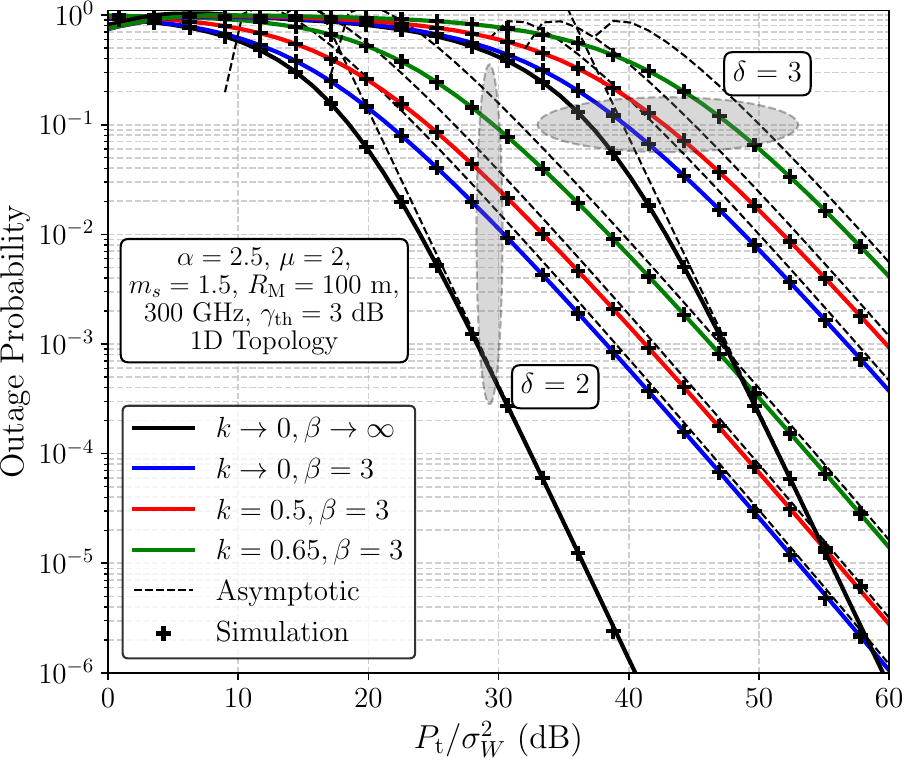}
\caption{OP as a function of $P_{\text{t}}/\sigma_{W}^{2}$ (in dB) for different path loss exponents ($\delta$) and hardware impairment levels ($k$). The benchmark case corresponds to ideal hardware ($k \rightarrow 0$) and perfect beam alignment ($\beta \rightarrow \infty$).}
\label{fig:op_result}
\end{figure}

\textcolor{black}{
Fig.~\ref{fig:op_result} presents the OP as a function of $P_{\text{t}}/\sigma_{W}^{2}$ (in dB) for different values of the path loss exponent and hardware impairment levels, considering an SNR threshold of $\gamma_{\text{th}}~=~3$~dB. The ideal case, defined by $k \rightarrow 0$ and $\beta \rightarrow \infty$ (\textit{i.e.}, no hardware impairments and perfect beam alignment), is adopted as a benchmark. The remaining parameters are set as $\alpha = 2.5$, $\mu = 2$, $m_{s} = 1.5$, and $f = 300$~GHz, assuming a 1D propagation environment with a maximum link distance of $R_{M} = 100$~m. The results highlight the pronounced impact of path loss on system performance. In particular, scenarios with $\delta = 3$ exhibit a significantly higher OP compared to the free-space case ($\delta = 2$). This marked degradation, even at high values of $P_{\text{t}}/\sigma_{W}^{2}$, reflects the severe propagation conditions inherent to THz-band communications. Hardware impairments, quantified by the parameter $k$, also play a critical role. For instance, at $P_{\text{t}}/\sigma_{W}^{2} = 30$~dB and $\delta = 2$, the OP for the ideal benchmark case is approximately $7 \times 10^{-3}$, which is about two orders of magnitude lower than the case with $k = 0.65$ and $\beta = 3$. Finally, the asymptotic curves show good agreement with the exact analytical results in the high-SNR regime, and the simulation results consistently validate the analytical expressions. 
As attested by our theoretical findings, the diversity order, which is represented by
the slope of asymptotic curves, depended only on the number of multipath clusters, and the channel non-linearity and pointing error parameters.}

\begin{figure}
\centering
\includegraphics[width=0.7\linewidth]{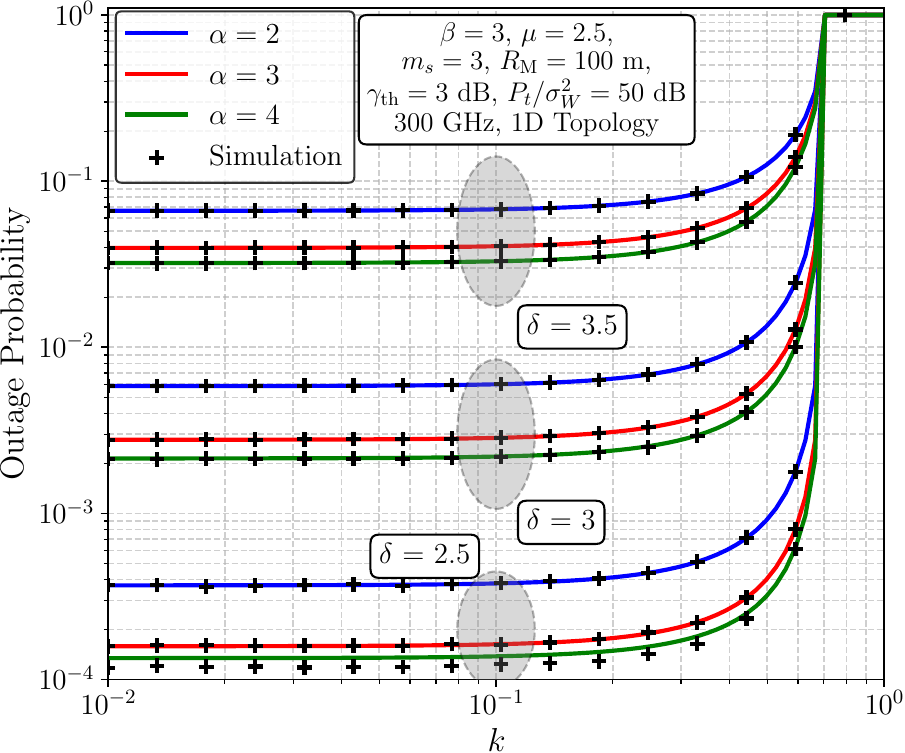}
\caption{OP as a function of the hardware impairment level ($k$) for different values of the nonlinearity factor ($\alpha$) and path loss exponent ($\delta$).}
\label{fig:op_result_b}
\end{figure}

\textcolor{black}{
Fig.~\ref{fig:op_result_b} depicts the OP as a function of the hardware impairment level $k$ for different values of the nonlinearity factor $\alpha$ and the path loss exponent $\delta$. The remaining parameters are set to $\beta = 3$, $\mu = 2.5$, $m_{s} = 3$, $f = 300$~GHz, $P_{\text{t}}/\sigma_{W}^{2} = 50$~dB, $\gamma_{\text{th}} = 3$~dB, and $R_{M} = 100$~m, assuming a 1D topology. The results further confirm the impact of path loss on system performance. In particular, more severe propagation conditions ($\delta = 3.5$) lead to a significant degradation in OP compared to milder scenarios ($\delta = 2.5$). In addition, a slight improvement in OP is observed as the nonlinearity factor $\alpha$ increases. Regarding hardware impairments, the system becomes less sensitive to $k$ for values below approximately $k = 0.1$ under the considered $P_{\text{t}}/\sigma_{W}^{2}$ ratio. Conversely, for sufficiently large values of $k$, the OP approaches unity for the adopted threshold. This behavior arises because hardware impairments impose an upper bound on the achievable SNR. Specifically, when $1/k^{2} < \gamma_{\text{th}}$ (i.e., $k \approx 0.7079$ for $\gamma_{\text{th}} = 3$ dB), the condition $\text{Prob}[\gamma < \gamma_{\text{th}}] = 1$ holds. Finally, the analytical results are in agreement with the Monte Carlo simulations, confirming the accuracy of the proposed model.}

\begin{figure}
\centering
\includegraphics[width=0.7\linewidth]{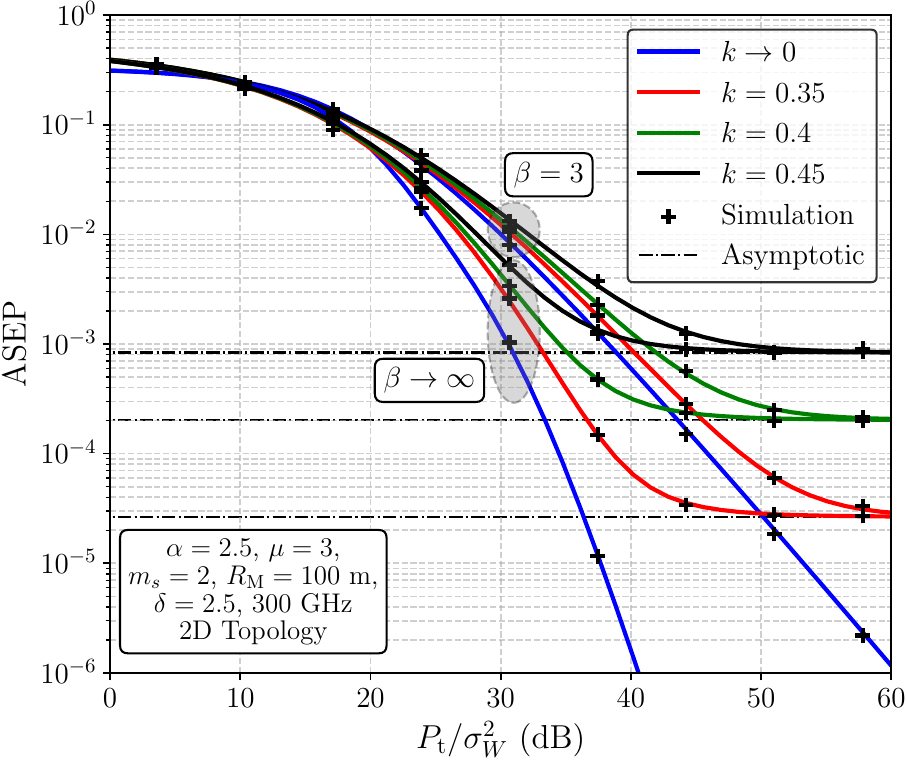}
\caption{ASEP as a function of $P_{\text{t}}/\sigma_{W}^{2}$ for different levels of hardware impairments ($k$) and beam misalignment ($\beta$), considering BPSK modulation ($a = 1$, $b = 2$). The ideal case ($k \rightarrow 0$, $\beta \rightarrow \infty$) is included as a benchmark.}
\label{fig:asep_result}
\end{figure}

\textcolor{black}{
Fig.~\ref{fig:asep_result} presents the ASEP as a function of $P_{\text{t}}/\sigma_{W}^{2}$ for different levels of hardware impairments and beam misalignment, including the ideal benchmark case defined by $\beta \rightarrow \infty$ and $k \rightarrow 0$. The analysis assumes a BPSK constellation ($a = 1$ and $b = 2$), with parameters $\alpha = 2.5$, $\mu = 3$, $m_{s} = 2$, $\delta = 2.5$, $f~=~300$~GHz, and $R_{M} = 100$~m, considering a 2D topology. The results demonstrate the significant impact of beam misalignment on ASEP performance. In particular, the ASEP can increase by more than one order of magnitude when comparing the case $\beta = 3$ with the ideal scenario ($\beta \rightarrow \infty$). Moreover, hardware impairments impose a lower bound on ASEP, leading to an irreducible error floor characterized by the constant $c_{3}$, as defined in~\eqref{eq:asep_hsnrlim}. This behavior occurs because, for sufficiently large values of $P_{\text{t}}/\sigma_{W}^{2}$, the SNR saturates at $1/k^{2}$, effectively reducing the channel to an AWGN channel with constant SNR. Consequently, the ASEP converges to $c_{3}$, and further increases in $P_{\text{t}}/\sigma_{W}^{2}$ do not yield performance gains. For relatively large values of $k$, this effect results in significant and irreducible error rates.}

\begin{figure}
\centering
\includegraphics[width=0.7\linewidth]{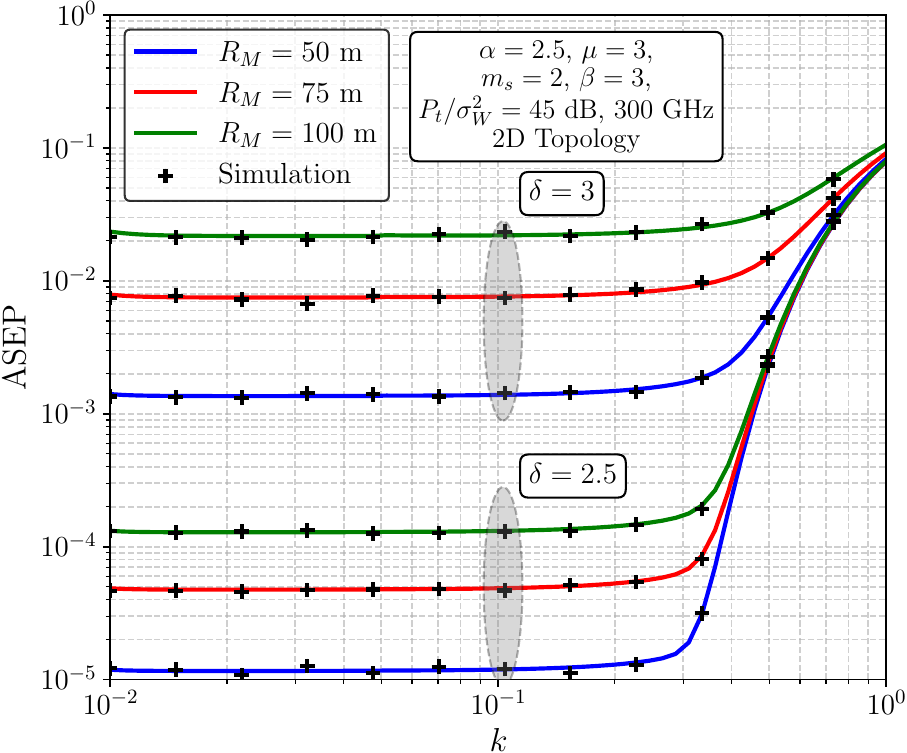}
\caption{ASEP as a function of the hardware impairment level $k$ for different path loss exponents $\delta$ and maximum link distances $R_{M}$.}
\label{fig:asep_result_b}
\end{figure}

\textcolor{black}{
Fig.~\ref{fig:asep_result_b} illustrates the ASEP as a function of the hardware impairment level $k$ for different values of the path loss exponent $\delta$ and maximum link distance $R_{M}$. The remaining parameters are set to $\alpha = 2.5$, $\mu = 3$, $m_{s} = 2$, $\beta = 3$, $P_{\text{t}}/\sigma_{W}^{2} = 45$~dB, and $f = 300$~GHz, considering a 2D topology. The results indicate that ASEP is highly sensitive to the combined effects of propagation distance and path loss. In particular, for low values of $k$, scenarios with $\delta = 3$ exhibit performance degradation of approximately two orders of magnitude compared to $\delta = 2.5$. Regarding hardware impairments, their impact becomes negligible for $k < 0.1$ under the considered $P_{\text{t}}/\sigma_{W}^{2}$ ratio. However, as $k$ increases, ASEP rises significantly across all evaluated scenarios, becoming a dominant limiting factor. For instance, ASEP reaches values on the order of $10^{-1}$ in all evaluated scenarios when $k = 1$. }

\begin{figure}
\centering
\includegraphics[width=0.75\linewidth]{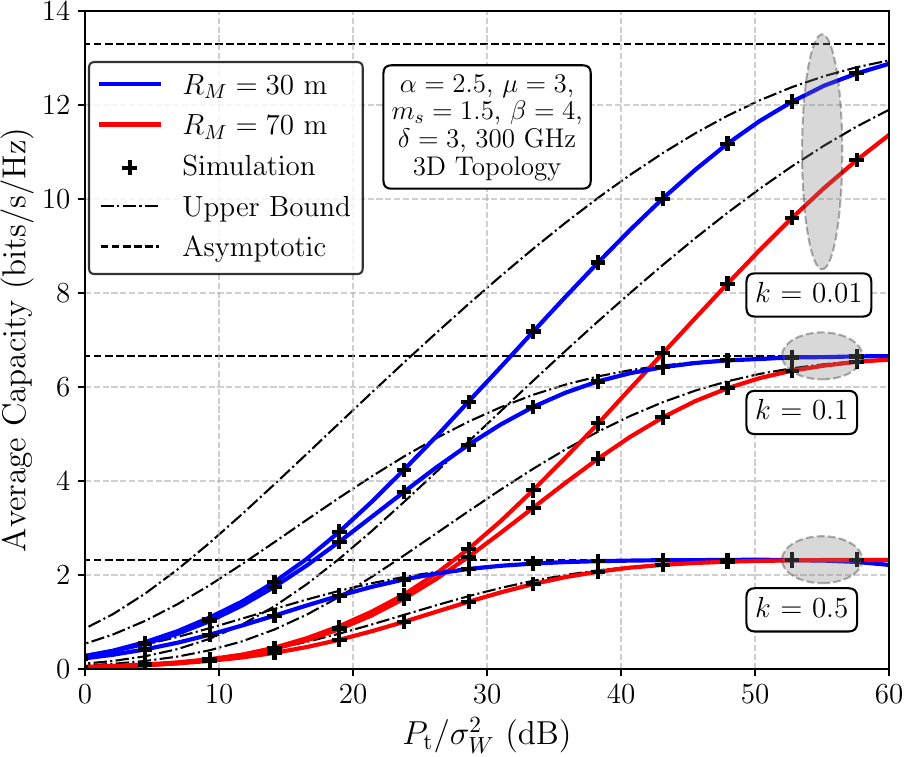}
\caption{Average capacity as a function of $P_{\text{t}}/\sigma_{W}^{2}$ for different maximum link distances ($R_{M}$) and hardware impairments leves ($k$).  }
\label{fig:cap_result}
\end{figure}

\textcolor{black}{
Fig.~\ref{fig:cap_result} presents the average capacity curves as a function of $P_{\text{t}}/\sigma_{W}^{2}$ for various maximum link distances $R_{M}$ and hardware impairment levels $k$, along with the corresponding upper limit curves for reference. The remaining parameters are fixed as $\alpha=2.5$, $\mu=3$, $m_{s}=1.5$, $\beta=4$, $\delta=3$, $f=300$~GHz, assuming a 3D topology. As expected, increasing the maximum link distance reduces the average capacity due to greater path loss and atmospheric absorption over longer links, a trend clearly observed when comparing $R_{M}=30$~m and $R_{M}=70$~m. Consistent with the observations for ASEP, hardware impairments impose an maximum limit on the average capacity. Under high SNR conditions (i.e., large $P_{\text{t}}/\sigma_{W}^{2}$), the capacity asymptotically approaches $\log_2(1+1/k^2)$, which cannot be exceeded. This limitation becomes particularly restrictive for large $k$. For instance, when $k=0.5$, the maximum achievable capacity is only 2.32~bits/s/Hz, whereas for $k=0.01$, it reaches 13.28~bits/s/Hz. These results highlight how the combined effects of distance-dependent losses and hardware impairments severely constrain link performance in the THz band. The upper bound curves closely follow the analytical trends, converging near $\log_2(1+1/k^2)$. Moreover, the simulation results show excellent agreement with the analytical curves, confirming the accuracy of the derived expressions.}

\begin{figure}
\centering
\includegraphics[width=0.7\linewidth]{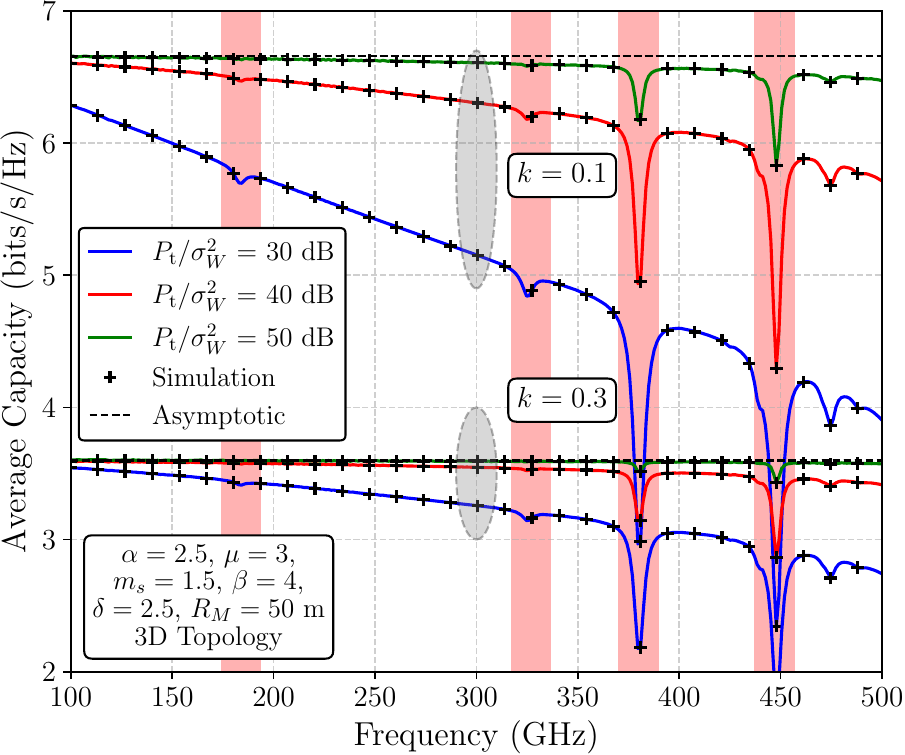}
\caption{Average capacity as a function of frequency for different values of $P_{\text{t}}/\sigma_{W}^{2}$ and hardware impairment levels ($k$). Frequency ranges with resonance peaks are highlighted.}
\label{fig:cap_result_b}
\end{figure}

\textcolor{black}{
As a final result, Fig.~\ref{fig:cap_result_b} presents the average capacity as a function of frequency for different values of $P_{\text{t}}/\sigma_{W}^{2}$ and hardware impairment levels $k$. The parameters are set to $\beta = 4$, $\alpha = 2.5$, $\mu = 3$, $m_{s} = 1.5$, $\delta = 2.5$, and $R_{M} = 50$~m, assuming a 3D propagation environment. The results show a general decreasing trend in average capacity with increasing frequency, mainly due to the higher path losses at THz bands. In addition, high-frequency communications are strongly impacted by atmospheric molecular absorption, modeled by the factor $\kappa(f)$. Specific frequency bands are particularly affected due to molecular resonance phenomena, resulting in pronounced absorption peaks around 184~GHz, 327~GHz, 380~GHz, and 447~GHz, which significantly degrade the average capacity. Furthermore, hardware impairments impose a maximum limit on the achievable capacity in the high-SNR regime, given by $\log_{2}(1 + 1/k^{2})$. For instance, this limit is approximately $3.6$~bits/s/Hz for $k = 0.3$ and $6.65$~bits/s/Hz for $k = 0.1$. It is observed that, for high SNR values (e.g., $P_{\text{t}}/\sigma_{W}^{2} = 50$~dB) and frequencies below 300~GHz, the capacity saturates and is upper limited by the hardware impairments. Conversely, for lower SNR levels and higher carrier frequencies, frequency-dependent effects dominate, leading to a reduction in average capacity as frequency increases.}

\section{Conclusions} \label{conclusao}

\textcolor{black}{ This article presented a comprehensive analysis of THz systems over $\alpha$-$\mathcal{F}$ fading channels, taking into account the joint effects of beam misalignment, user mobility, and transceiver hardware impairments.
New closed-form expressions were derived for key statistical functions, including the PDF, CDF, and higher-order moments of the SNR. Additionally, novel formulations for OP, ASEP, and channel capacity were obtained. Asymptotic expressions were also developed to provide deeper insights into the behavior of these metrics under varying channel conditions and system parameters. An upper bound on the capacity metric was also obtained.}

\textcolor{black}{The results revealed that THz system performance is significantly influenced by the interplay among channel nonlinearity, shadowing, and multipath fading, as captured by the $\alpha$-$\mathcal{F}$ model. Furthermore, beam misalignment and mobility were shown to introduce substantial performance degradation. Hardware impairments emerged as a critical limiting factor, imposing a strict upper bound on the achievable SNR, which leads to saturation effects in both error performance and capacity. In particular, the presence of imperfections is reflected in irreducible error and capacity limits, even in the high SNR regime. The analytical findings were validated through Monte Carlo simulations, demonstrating excellent agreement and confirming the accuracy of the proposed framework.}

\textcolor{black}{For future works, several extensions can be considered. The incorporation of physical layer security metrics is a promising research avenue, especially under the combined effects of fading, misalignment, mobility, and hardware impairments. Moreover, the proposed framework can be extended to multi-user scenarios, where multiple antennas and antenna correlation effects should be comprehensively considered.
}


\appendices

\section{}\label{FirstAppendix}
\textcolor{black}{From (\ref{eq:snr}), the PDF of $\Gamma(d)$ given $d$ can be written as
\begin{align}
f_{\Gamma}(\gamma|d)&=\frac{\partial}{\partial\gamma}\text{Prob}[\Gamma(d)\leq\gamma] \nonumber \\
                    &=\frac{\partial}{\partial\gamma}
\text{Prob}\left[V\leq\frac{(d-1)^{\delta}\exp[\kappa(f)d_{0}d]}{\gamma_{0}(1-k^{2}\gamma)}\gamma\right] \nonumber \\
&=\frac{(d-1)^{\delta}\exp[\kappa(f)d_{0}d]}{\gamma_{0}(1-k^{2}\gamma)^{2}} \nonumber \\
&\times f_{V}\left((d-1)^{\delta}\frac{\exp[\kappa(f)d_{0}d]\gamma}{\gamma_{0}(1-k^{2}\gamma)}\right),
\end{align}    
in which the PDF of the instantaneous power $V=XZ=H_{\text{f}}^{2}H_{\text{p}}^{2}$ can be derived substituting
(\ref{eq:fundZz}) and (\ref{eq:fundXx}) into (\ref{eq:fundVv}), 
\begin{align}
f_{V}(v)=\int_{-\infty}^{\infty}\frac{1}{|y|}f_{X}\left(\frac{v}{y}\right)f_{Z}(y)\text{d}y.
\label{eq:fundVv}
\end{align}
}

\textcolor{black}{We assume that $\mathbb{E}[V]=\mathbb{E}[H_{\text{f}}^{2}H_{\text{p}}^{2}]=1$, so $f_{V}(v)$ can be written as
\begin{align}
f_{V}(v)=c_{1}v^{\frac{\alpha \mu}{2}-1}\int_{0}^{\infty}\frac{ye^{-\frac{y}{2}(\beta - \alpha \mu)}}
{\left(v^{\frac{\alpha}{2}}e^{\frac{y\alpha}{2}}+c_{2}\right)^{(m_{s}+\mu)}}\text{d}y,
\label{fundVv2}
\end{align}
in which
\begin{equation}\label{eq:Constantec1}
c_{1}=\frac{\beta^{2}\alpha}{8 \text{B}(\mu,m_{s})} \left(\frac{(m_{s}-1)(\beta + 2)^\alpha }{\beta^\alpha \lambda^{\frac{\alpha}{2}} \mu }\right)^{m_{s}} 
\end{equation}
and
\begin{equation}\label{eq:Constantec2}
c_{2}=\frac{(m_{s}-1)(\beta + 2)^\alpha}{ \lambda^{\frac{\alpha}{2}}\mu \beta^{\alpha} }.    
\end{equation}
}

\textcolor{black}{
Since the Mellin transform of $(1+z)^{-a}$ is $\text{B}(\zeta,a-\zeta)$, then $(1+z)^{-a}$ can be written in terms of the 
inverse Mellin transform as
\begin{equation}
(1+z)^{-a}=\frac{1}{j2\pi\Gamma(a)}\int_{\mathcal{L}}\Gamma(\zeta)\Gamma(a-\zeta)z^{-\zeta}\text{d}\zeta.
\label{binomza}
\end{equation}
Expressing the binomial term of  
(\ref{fundVv2}) like in (\ref{binomza}) and the Mellin-Barnes integral in terms of the Fox H-function, one can 
write the conditional PDF $f_{\Gamma}(\gamma|d)$ as
\begin{align}
&f_{\Gamma}(\gamma|d)=\frac{4c_{1}c_{2}^{-(m_{s}+\mu)}}{\Gamma(m_{s}+\mu)\gamma_{0}^{\frac{\alpha\mu}{2}}}
\frac{\gamma^{\frac{\alpha\mu}{2}-1}}{\left(1-k^{2}\gamma\right)^{\frac{\alpha\mu}{2}+1}}\nonumber\\
&\times \frac{1}{j2\pi}\int_{\mathcal{L}}\phi(\zeta)\left[\frac{1}{c_{2}\gamma_{0}^{\frac{\alpha}{2}}}
\left(\frac{\gamma}{1-k^{2}\gamma}\right)^{\frac{\alpha}{2}}\right]^{-\zeta}
(d-1)^{\frac{\alpha\delta}{2}(\mu-\zeta)}\nonumber\\
&\times\exp\left[\kappa(f)d_{0}d\left(\frac{\alpha\mu}{2}-\frac{\alpha\zeta}{2}\right)\right]\mathrm{d}\zeta,
\end{align}
in which
\begin{equation}
\phi(\zeta)=\frac{\Gamma(\zeta) \Gamma^2(\beta - \alpha \mu +\alpha \zeta) \Gamma(\mu+m_{s} -\zeta)}
{\Gamma^2(1 + \beta - \alpha \mu +\alpha \zeta)}.
\label{eq2:Fapendix}    
\end{equation}
}

\textcolor{black}{
The PDF $f_{\Gamma}(\gamma)$ can be derived as
\begin{equation}
f_{\Gamma}(\gamma)=\int_{0}^{\infty} f_{\Gamma}(\gamma|d) f_{D}(d) \text{d}d,
\end{equation}
in which $f_{D}(d)$ is given by (\ref{eq:fDundd}). Therefore, $f_{\Gamma}(\gamma)$ is
\begin{align}
&f_{\Gamma}(\gamma)=\frac{4c_{1}c_{2}^{-(m_{s}+\mu)}}{\Gamma(m_{s}+\mu)\gamma_{0}^{\frac{\alpha\mu}{2}}}\frac{d_{0}}{R_{M}}
\frac{\gamma^{\frac{\alpha\mu}{2}-1}}{\left(1-k^{2}\gamma\right)^{\frac{\alpha\mu}{2}+1}}\nonumber\\
&\times\frac{1}{j2\pi}\int_{\mathcal{L}}\phi(\zeta)\left[\frac{1}{c_{2}\gamma_{0}^{\frac{\alpha}{2}}}
\left(\frac{\gamma}{1-k^{2}\gamma}\right)^{\frac{\alpha}{2}}\right]^{-\zeta}\sum_{i=1}^{l}B_{i}
\left(\frac{d_{0}}{R_{M}}\right)^{\beta_{i}}\nonumber\\
&\times\int_{1}^{1+\frac{R_{M}}{d_{0}}}\frac{(d-1)^{\frac{\delta\alpha\mu}{2}+\beta_{i}-\frac{\delta\alpha}{2}\zeta}}
{\exp\left[-\kappa(f)d_{0}d\frac{\alpha}{2}(\mu-\zeta)\right]}\mathrm{d}d\mathrm{d}\zeta.
\label{eqfprp}
\end{align}
Transforming the interval $[1,1+R_{M}/d_{0}]$ of (\ref{eqfprp}) to $[-1,1]$ by means of 
\cite[Eq. (3.021.1)]{Gradsteyn2007}, defining the auxiliary function
\begin{equation}
\Phi(x;\mathbf{a}^{T})=[a_{1}(1+x)^{a_{2}}]\exp[a_{3}(a_{1}(1+x)+1)]    
\end{equation}
and considering the Gauss-Legendre quadrature, $f_{\Gamma}(\gamma)$ can be written as
\begin{align}
&f_{\Gamma}(\gamma)=\frac{2R_{M}c_{1}c_{2}^{-(m_{s}+\mu)}}{d_{0}\Gamma(m_{s}+\mu)
\gamma_{0}^{\frac{\alpha\mu}{2}}}
\frac{\gamma^{\frac{\alpha\mu}{2}-1}}{\left(1-k^{2}\gamma\right)^{\frac{\alpha\mu}{2}+1}}\nonumber \\
&\times\sum_{i=1}^{l}\sum_{k=1}^{N}B_{i}\left(\frac{d_{0}}{R_{M}}\right)^{\beta_{i}+1}w_{k}\Phi(x_{k};\mathbf{v}_{1}^{T}(i))
\nonumber \\
&\times\frac{1}{j2\pi}\int_{\mathcal{L}}\phi(\zeta)\left[\frac{\Phi(x_{k};\mathbf{v}_{2}^{T})}{c_{2}\gamma_{0}^{\frac{\alpha}{2}}}
\left(\frac{\gamma}{1-k^{2}\gamma}\right)^{\frac{\alpha}{2}}\right]^{-\zeta}\mathrm{d}\zeta.
\label{eq:eqmelinbafrp}
\end{align}
Writing (\ref{eq:eqmelinbafrp}) in terms of the Fox H-function~\cite[Eq. (1.2)]{Mathai}, one gets the PDF $f_{\Gamma}(\gamma)$ shown in~(\ref{eq:mobilityPDF}). 
}

\textcolor{black}{The CDF of $\Gamma$, shown in (\ref{eq:mobilityCDF}), is derived replacing 
(\ref{eq:eqmelinbafrp}) in 
\begin{equation}
F_{\Gamma}(\gamma)=\left\{\begin{array}{ll}
\int_{0}^{\gamma}f_{\Gamma}(v)\text{d}v, & \gamma < 1/k^{2} \\
1, & \gamma\geq 1/k^2
\end{array}\right..
\end{equation}
mapping the integration interval from $[0,\gamma]$ to $[-1,1]$ by means of \cite[Eq. (3.021.1)]{Gradsteyn2007}, applying 
the Gauss-Legendre quadrature and writing the result in terms of the Fox H-function. Thus, the proof is complete.
}

\section{}\label{SecondAppendix}
\textcolor{black}{The $n$-th moment of the instantaneous SNR $\Gamma$, denoted by  $\mathbb{E}[\Gamma^{n}]$, is calculated as
\begin{equation}\label{eq:defExpec}
\mathbb{E}[\Gamma^{n}]=\int_{0}^{\infty}\gamma^{n}f_{\Gamma}(\gamma)\mathrm{d}\gamma.    
\end{equation}
Replacing (\ref{eq:eqmelinbafrp}) in~(\ref{eq:defExpec}), it follows that
\begin{align}
&\mathbb{E}[\Gamma^{n}]=\frac{2R_{M}c_{1}c_{2}^{-(\mu+m)}}{d_{0}\Gamma(\mu+m)\gamma_{0}^{\frac{\alpha\mu}{2}}}
\sum_{i=1}^{l}\sum_{k=1}^{N}B_{i}\left(\frac{d_{0}}{R_{M}}\right)^{\beta_{i}+1} \nonumber \\
&\times w_{k}\Phi(x_{k};\mathbf{v}_{1}^{T}(i))\int_{\mathcal{L}}\phi(\zeta)\left(\frac{\Phi(x_{k};\mathbf{v}_{2}^{T})}
{c_{2}\gamma_{0}^{\frac{\alpha}{2}}}\right)^{-\zeta}\mathcal{I}(\zeta)\mathrm{d}\zeta,
\end{align}
in which
$
\mathcal{I}(\zeta)=\int_{0}^{\frac{1}{k^{2}}}\gamma^{n+\frac{\alpha\mu}{2}-1-\frac{\alpha}{2}\zeta}
(1-k^{2}\gamma)^{\frac{\alpha}{2}\zeta-\frac{\alpha\mu}{2}-1}\mathrm{d}\gamma.    
$
Expressing $\mathcal{I}(\zeta)$ in terms of the Beta function, writing the Beta function as a ratio of Gamma functions and writing 
the resulting Mellin-Barnes integral in terms of the Fox H-function, one gets the result in (\ref{eq:rawmoment}). This complete the proof.
}

\balance


\begin{thebibliography}{}

\bibitem{Thomas2025} S. Thomas et al, ``A Survey on Advancements in THz Technology 
for 6G: Systems, Circuits, Antennas, and Experiments'',\textit{IEEE Open J. Commun. Soc.}, vol. 6, pp. 1998-2016, Mar. 2025.


\bibitem{Dabiri} M. T. Dabiri and M. Hasna, ``Pointing Error Modeling of mmWave to THz High-Directional Antenna Arrays,'' \textit{IEEE Wirel. Commun. Lett.}, vol. 11, no. 11, pp. 2435-2439, Nov. 2022.

\bibitem{Chen} W. Chen et al, ``Mobility and Blockage-induced Beam Misalignment and Throughput Analysis for THz Networks,'' in Proc. of the \textit{IEEE Global Communications Conference}, 2021.


\bibitem{Pattaramalai} E. Meesa-Ard and S. Pattaramalai, ``Evaluating the Mobility Impact on the Performance of Heterogeneous Wireless Networks Over $\eta$-$\mu$ Fading Channels,'' \textit{IEEE Access}, vol. 9, pp. 65017-65032, Apr. 2021.


\bibitem{Shawaqfeh} M K. Shawaqfeh and O. S. Badarneh, ``Performance of Mobile Networks under Composite Fading Channels'', \textit{Digit. Commun. Netw.}, vol. 8, no. 1, pp. 25-32, Feb. 2022. 

\bibitem{OsamahPoint} O. S. Badarneh, M. T. Dabiri and M. Hasna, ``Channel Modeling and Performance Analysis of Directional THz Links Under Pointing Errors and $\alpha$-$\mu$ Distribution,'' \textit{IEEE Commun. Lett. }, vol. 27, no. 3, pp. 812-816, Mar. 2023.

\bibitem{Pereira} P. M. R. Pereira et al, ``Mobility, Path Loss, and Composite Fading: Performance of a Conventional and of a Non-Conventional System With a Robust Autoencoder,'' \textit{IEEE Trans. Veh. Technol.}, vol. 72, no. 12, pp. 16725-16730, Dec. 2023.


\bibitem{Jemaa} H. Jemaa et al., ``Performance Analysis of Outdoor THz Links Under Mixture Gamma Fading With Misalignment,'' \textit{IEEE Commun. Lett. }, vol. 28, no. 11, pp. 2668-2672, Nov. 2024.


\bibitem{Bjornson} E. Bjornson, M. Matthaiou and M. Debbah, ``A New Look at Dual-hop Relaying: Performance Limits with Hardware Impairments'' \textit{IEEE Trans. Commun.}, vol. 61, no. 11, pp. 4512-4525, Nov. 2013.


\bibitem{Boulogeorgos} A. A. Boulogeorgos, E. N. Papasotiriou and A. Alexiou, ``Analytical Performance Assessment of THz Wireless Systems,'' \textit{IEEE Access}, vol. 7, pp. 11436-11453, Jan. 2019.

\bibitem{Bhardwaj} P. Bhardwaj and S. M. Zafaruddin, ``Exact Performance Analysis of THz Link Under Transceiver Hardware Impairments,'' \textit{IEEE Commun. Lett.}, vol. 27, no. 8, pp. 2197-2201, Aug. 2023.

\bibitem{AlBadarneh} Y. H. Al-Badarneh et al., ``Capacity of Wireless Channels under Transceiver Hardware Impairments and Adaptive Transmission Techniques'' \textit{IEEE Commun. Lett.}, vol. 29, no. 4, pp. 764-768, Apr. 2025.

\bibitem{Badarneh} O. S. Badarneh, ``The $\alpha$-$\mathcal{F}$ Composite Fading Distribution: Statistical Characterization and Applications,''  \textit{IEEE Trans. Veh. Technol.}, vol. 69, no. 8, pp. 8097-8106, Aug. 2020.

\bibitem{Almeida2023} P. H. D. Almeida et al. The $\alpha$-${\cal{F}}$ Composite Distribution with Pointing Errors: Theory and Applications to RIS. \textit{J. Frank. Inst.}, vol. 361, no. 10, pp. 1-13, Jul. 2024.


\bibitem{Wolfram} Wolfram Research, Inc. (2020). \textit{Wolfram Research}. Accessed: May. 5, 2025. [Online]. Available: http://functions.wolfram.com/id.


\bibitem{Mathai} A. M. Mathai, R. K. Saxena, and H. J. Haubold. \textit{The H-Function: Theory and Applications}, 1st edition, New York: Springer, 2009.



\bibitem{Papasotiriou2021b} E. N. Papasotiriou et al., ``An Experimentally Validated Fading Model for THz Wireless Systems,'' \textit{Scientific Reports}, vol. 11, no. 18717, pp. 1-14, Sep. 2021.

\bibitem{ITU} International Telecommunication Union, ``Recommendation ITU‑R P.676‑13: Attenuation by atmospheric gases and related effects,'' Radiowave Propagation Series P, approved Aug. 2022, electronic publication, Geneva, 2022. [Online]. Available: https://www.itu.int/rec/R-REC-P.676-13-202208-I/en




\bibitem{HTransforms} A. A. Kilbas and M. Saigo. \textit{H-Transforms: Theory and Applications}. Florida: CRC Press Inc, 2004.

\bibitem{rao2015mgf} M. Rao et al, ``MGF Approach to the Capacity Analysis of Generalized Two-Ray Fading Models'', in Proc. of the \textit{IEEE International Conference on Communications (ICC)}, 2015.


\bibitem{Weisstein} E. W. Weisstein. \textit{Digamma Function, from MathWorld - A Wolfram Resource}. Accessed: June 22, 2025. 
[Online]. Available: https://mathworld.wolfram.com/DigammaFunction.html.  

\bibitem{Gradsteyn2007}
I. S. Gradshteyn and  I. M. Ryshik. \textit{Table of integrals, series, and products}. California: Elsevier Inc, 2007. 

\bibitem{Rappaport2019}
T. S. Rappaport et al., ``Wireless Communications and Applications Above 100 GHz: Opportunities and Challenges for 6G and Beyond'', \textit{IEEE Access}, vol. 7, pp. 78729-78757, Jun. 2019.






















\end{thebibliography}
\end{document}